\documentclass[11pt]{article}

\usepackage[margin=1in]{geometry}
\usepackage[T1]{fontenc}
\usepackage[utf8]{inputenc}
\usepackage{lmodern}
\usepackage{microtype}
\usepackage{booktabs,tabularx,array,multirow}
\usepackage{graphicx}
\usepackage{xcolor}
\usepackage{enumitem}
\usepackage{amssymb}
\usepackage{hyperref}
\usepackage[nameinlink,noabbrev]{cleveref}
\usepackage[numbers,sort&compress]{natbib}
\usepackage{tikz}
\usetikzlibrary{arrows.meta,positioning,shapes.geometric,fit,backgrounds,calc}
\hypersetup{colorlinks=true,citecolor=blue!60!black,linkcolor=blue!60!black,urlcolor=blue!60!black}
\newcolumntype{Y}{>{\raggedright\arraybackslash}X}

\title{Experimental Settings in LLM-Based Program Repair: \\ A Study of Inputs, Tool Access, Feedback, and Validation}
\author{Xushu Dai \quad Yicheng Cai \quad Nanqing Luo \quad Pei-Yu Tseng\\[0.5em]
Pennsylvania State University\\
\texttt{\{xsdai,yccai,nqluo,pmt5342\}@psu.edu}}
\date{September 2026}

\begin{document}
\maketitle

\begin{abstract}
Evaluations of automated program repair (APR) systems commonly report the benchmark, the number of repaired defects, and the tests used for final patch validation, but these items no longer fully specify the repair task presented to a system. Recent LLM-based systems differ in the information supplied before repair, the repository and testing operations permitted during repair, and the feedback returned after unsuccessful attempts, allowing the same benchmark to instantiate substantially different repair tasks ranging from localized patch generation to repository-level diagnosis and iterative repair. We present a framework for explicitly specifying the experimental settings associated with reported APR results. We analyze reported experimental settings from systems evaluated on Defects4J and SWE-bench and characterize each result by its task unit, fault-localization assumptions, initial input, tool access, repair-time feedback, final validation, and resource budget. Our analysis shows that benchmark identity alone is insufficient to reconstruct the evaluated task or determine the appropriate scope of comparison across reported repair rates. We therefore introduce a machine-readable schema for specifying each experimental setting to improve reproducibility and make the scope of cross-system comparisons explicit.
\end{abstract}

\section{Introduction}
\label{sec:intro}

Automated program repair (APR) and closely related repository-level issue-resolution tasks aim to generate source-code changes that address a reported failure or issue without introducing regressions. A conventional repair pipeline includes fault localization, patch generation, and patch validation~\citep{genprog,prophet,semfix,angelix,learningbasedsurvey}. Accordingly, an evaluation must specify what information about the issue, observed failure, and program is provided to the repair system, whether the relevant location is supplied or must be identified, and how candidate patches are validated.

Compared with many conventional APR settings, which prescribe a code region or a fixed repair pipeline, recent LLM-based and agentic systems can access a substantially broader set of information and operations. Recent methods use bug reports, compiler messages, failing tests, retrieved examples, and static-analysis results~\citep{chatrepair,inferfix,maniple,thinkrepair}, while repository-level systems may search files, inspect history, execute builds and tests, invoke analysis tools, and use command output to guide subsequent edits~\citep{sweagent,agentless,autocoderover,repairagent}. We distinguish between an operation that a system is permitted to invoke and the information returned by that operation: for example, permission to run tests is part of \emph{tool access}, whereas test outcomes, error messages, and stack traces are \emph{repair-time feedback}. These distinctions are especially important for agentic repair systems, where the information available during repair depends not only on the initial context but also on the operations it is permitted to perform.

APR evaluations are commonly organized and compared by benchmark, implicitly treating a shared benchmark as defining a sufficiently similar repair task. A benchmark provides a collection of program and evaluation artifacts rather than a unique interaction protocol. Studies using the same benchmark may expose different inputs, assign different portions of the repair workflow to the system, and provide different operations and feedback during repair.

For example, Defects4J can be used in an evaluation in which an oracle provides the faulty method and the system only generates a patch, as well as in an evaluation in which the system begins from a project checkout and failure information and must first locate the fault. Likewise, tests may be reserved exclusively for final validation, or their failures may be returned after each unsuccessful candidate and used to guide subsequent repair attempts. Both evaluation designs address meaningful but different repair settings: they assign different responsibilities to the repair system and expose different information during repair.

The resulting problem is under-specification of the experimental setting associated with a repair rate. From a benchmark name and score alone, a reader cannot reliably reconstruct the system's available observations, permitted actions, or division of responsibility. This limits faithful reproduction, obscures whether results obtained under the same benchmark are directly comparable, and weakens interpretations that attribute score differences to a model or repair method. A controlled ablation can isolate one factor within a single implementation. It does not recover the conditions of independently reported settings or provide a common account of information exposure, tool access, feedback, and final validation across results. We address this problem with a result-level \emph{repair evaluation protocol} and machine-readable schema. The protocol separates benchmark artifacts, initial input, permitted actions and runtime observations, and final validation. It enables readers to reconstruct the task contract, identify materially different settings built from the same benchmark, and determine which comparisons a reported result can support.

Prior APR work records fault-localization assumptions, input information, benchmark variants, and evaluation settings~\citep{learningbasedsurvey,llmaprslr,llmaprsurvey,patchgensurvey}. We use \emph{experimental setting} (or \emph{record}) for the conditions associated with one reported result. Rather than introducing these factors individually, we provide a result-level representation that encodes them jointly and distinguishes when each source of information becomes available: whether an artifact is stored, initially exposed, observed at runtime through a permitted action, or reserved for final validation. An experimental setting records task unit (e.g., method-, defect-, or repository-level), fault-localization setting, initial input, tool access, repair-time feedback, final validation, and resource budget. \emph{Configuration} is only the within-study label used to identify an experimental setting, such as perfect- and statistical-fault-localization variants of one system.

Our contributions are:
\begin{enumerate}[leftmargin=*,nosep]
\item We analyze a corpus of 10 experimental settings from seven studies, each evaluating one focal APR system on Defects4J or SWE-bench.
\item We introduce a machine-readable schema that records key evaluation factors, including fault localization, input information, tool access, repair-time feedback, final validation, and repair budget.
\item We demonstrate, within the studied corpus, that the same benchmark is used to instantiate materially different repair tasks, and identify experimental details that could not be recovered from the reported studies.
\item We provide reporting guidance for determining when repair rates support comparison at the complete-system level and when claims about a model or repair method require additional experimental controls.
\end{enumerate}

\begin{figure}[t]
\centering
\begin{tikzpicture}[font=\small, node distance=6mm and 12mm,
box/.style={draw, rounded corners=2pt, align=center, inner sep=4pt},
asset/.style={box, fill=gray!13, text width=112mm, minimum height=13mm},
gate/.style={box, fill=blue!12, text width=66mm, minimum height=12mm},
initial/.style={box, fill=blue!8, text width=33mm, minimum height=24mm},
agent/.style={box, fill=green!12, text width=33mm, minimum height=24mm},
oracle/.style={box, fill=orange!18, text width=33mm, minimum height=24mm},
arrow/.style={-{Stealth[length=2mm]}, thick, draw=black!75}]
\node[asset] (assets) {\textbf{Benchmark artifact pool}\quad issue / bug report \enspace $\cdot$ \enspace repository and history \enspace $\cdot$ \enspace tests and reproducer \enspace $\cdot$ \enspace gold patch};
\node[gate, below=of assets] (contract) {\textbf{Evaluation protocol $\mathcal{T}$}\quad specifies environment, initial input, permitted actions, feedback channels, and final oracle};
\node[agent, below=9mm of contract] (tools) {\textbf{Runtime observations}\\[1mm] repository search, build, test, analysis, debugger};
\node[initial, left=of tools] (initial) {\textbf{Initial input/context}\\[1mm] prompt, workspace, localization cue};
\node[oracle, right=of tools] (eval) {\textbf{Evaluation-only oracle}\\[1mm] held-out tests, reserved PoC, independent review};
\draw[arrow] (assets) -- (contract);
\draw[arrow] (contract.south west) -| (initial.north);
\draw[arrow] (contract.south) -- (tools.north);
\draw[arrow] (contract.south east) -| (eval.north);
\end{tikzpicture}
\caption{A benchmark provides program and evaluation artifacts. An experimental setting determines which artifacts and operations are available during repair and which tests remain reserved for final evaluation.}
\label{fig:layers}
\end{figure}
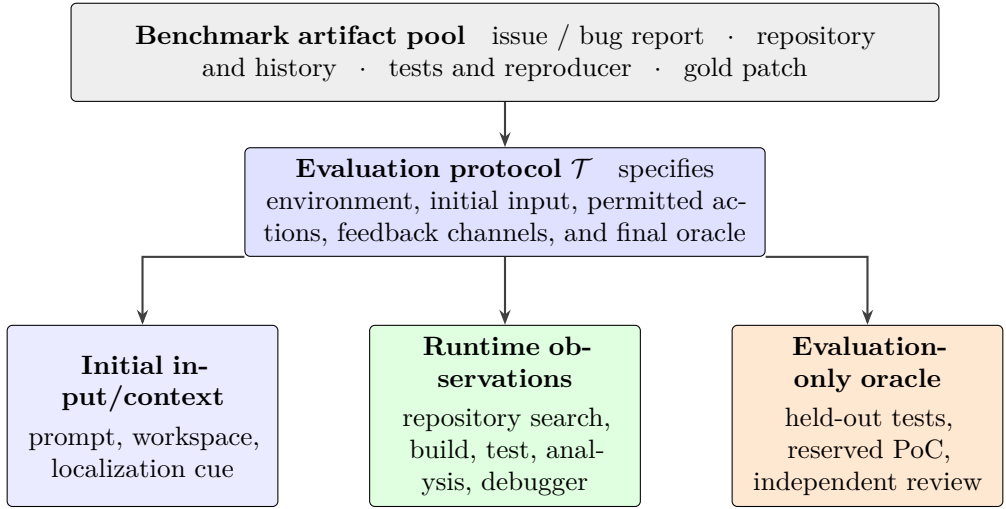

\section{Study design}
\label{sec:method}

The empirical scope covers source-code editing in response to a reported failure or repair-oriented repository issue. It includes test-suite-based APR, LLM patch generation, conversational repair, and repository-level issue-resolution agents, including SWE-bench instances whose issue reports may describe repository issues rather than observed failures. It excludes pure bug detection, decompilation, and unconstrained code generation not grounded in a reported failure or repository issue. The corpus consists of the seven studies and 10 experimental settings defined below, with sources accessed by 8 September 2026.

\subsection{Research questions and corpus scope}

We ask three descriptive research questions.
\begin{description}[leftmargin=*,style=nextline]
  \item[RQ1] How are APR evaluations set up in the selected studies?
  \item[RQ2] How do repair tasks differ across evaluations that use the same benchmark?
  \item[RQ3] Which experimental-setting details are reported clearly, and which remain missing or unclear?
\end{description}

We constructed a bounded corpus using maximum-variation sampling over repair-task designs. Our goal was to capture variation in experimental settings, not to estimate the prevalence of practices across APR. The sampling frame consisted of studies evaluating an LLM-based or agentic repair system on Defects4J or SWE-bench. We selected studies to span localized patch generation, iterative repair with feedback, and repository-level agents. Each study also had to report either a primary setting or a material setting variation and provide sufficient detail to code localization, initial evidence, and final validation. The corpus contains 10 records: two AlphaRepair localization settings, two ChatRepair repair scenarios, two ThinkRepair feedback settings, and one primary setting each for RepairAgent, Agentless, AutoCodeRover, and SWE-agent. A record is a tuple of study, benchmark, and configuration, not an individual run or bug. The corpus is intended to demonstrate and measure configuration variation within this defined sample. Its percentages are not prevalence estimates for APR. Because eligibility required a minimum level of reportable detail, the RQ3 missingness counts should not be interpreted as estimates for studies that fell below this coding threshold.

\subsection{Coding procedure and reproducibility}

For each study, we coded the primary reported description of its experimental setting. We applied a conservative rule: a benchmark artifact is not treated as model-visible unless the study establishes that it is included in the prompt or workspace, or returned by a permitted operation. We code missing information as \texttt{unclear}, never as an inferred default. A record represents a separately reported setting. We retain a separate record for a named setting variation or for an independently reported result, but do not create separate records solely because individual coded fields differ. The schema and coding rules support the descriptive counts reported in \Cref{sec:results}.

\subsection{Task-difference analysis}

For RQ2, we compare pairs of records evaluated on the same benchmark family. Independently of record creation, two records instantiate different repair tasks when their task units differ or when the model-visible or responsibility-defining parts of $\mathcal{T}$ differ: initial exposure $I$, localization regime $L$, permitted action interface $A$, runtime observation policy $\Omega$, or final oracle $V$. Differences only in system configuration $\mathcal{S}$, such as model version or decoding policy, do not by themselves establish different tasks. This rule makes RQ2 a comparison of reported record pairs, rather than a consequence of how the corpus was partitioned.

\section{Related work}
\label{sec:related}

Prior APR work records benchmark variants, metrics, base models, fault-localization assumptions, input/context, repository-level repair, leakage, and test-time scaling~\citep{llmaprsurvey,patchgensurvey}. These factors are therefore individually well recognized in the APR literature. Our contribution is a result-level encoding of their joint evidence flow: it separates an artifact's presence from its exposure and distinguishes initial exposure, runtime observations, and final validation.

Several adjacent lines of work help clarify the boundary of our contribution. MANIPLE studies which candidate bug-related facts should be selected for an LLM prompt~\citep{maniple}. Here, the available facts and their timing are part of the evaluation specification before fact selection begins. EviACT uses execution evidence to control retrieval, compilation, and test-driven decisions within an agentic repair method~\citep{eviact}. This article instead describes what evidence a reported system was permitted to access and what comparisons a resulting score can support. Beyond APR, the Scaffold Effect shows that a coding-agent harness can confound comparisons made by model name alone~\citep{scaffoldeffect}. Our framework specializes this broader concern to APR by recording the benchmark-to-evidence mapping, fault localization, action interface, feedback, and final validation oracle.

\section{From a benchmark to an experimental repair task}
\label{sec:contract}

\subsection{Four evaluation dimensions}
\label{sec:layers}

Figure~\ref{fig:layers} separates four dimensions of an APR experimental setting that must be distinguished when a benchmark is used to construct a repair task.

\paragraph{Benchmark artifacts.} These are artifacts contained in a dataset or evaluation harness: buggy and fixed revisions, issue text, test suites, tests labeled \emph{FAIL\_TO\_PASS}, reproduction scripts, exploit inputs, environment metadata, and gold patches. Their presence does not imply model visibility.

\paragraph{Initial input and context.} This is the information supplied before the first repair action. Examples include a localized function plus compiler error and an issue plus a repository checkout. Prompt contents and workspace contents both matter. An agent given a checkout can read an unmentioned test file.

\paragraph{Runtime observations and feedback.} An agent may obtain observations through permitted actions: searching code, reading history, running a build or test, invoking a static analyzer, or using a debugger. Tool output and test feedback are evaluation channels, not merely implementation details. Their availability, arguments, timeouts, and cost shape the task.

\paragraph{Evaluator-only oracles.} Held-out regression tests, a proof-of-concept reserved for validation, the developer patch, and manual correctness review are useful evaluation instruments. They should not be silently counted as diagnostic input. In particular, saying that a benchmark ``has a PoC'' does not establish that the model received that PoC.

\subsection{Repair evaluation protocol}
\label{sec:definition}

For a benchmark instance $b$, define a repair \emph{evaluation protocol} by
\[
\mathcal{T}=(E,I,L,A,\Omega,V),
\]
where $E$ is the environment state and transition semantics, including the workspace, dependencies, credentials, network policy, and artifacts whether or not the system may read them. $I$ is the model-visible initial input/context, and $L$ is the fault-localization regime and any supplied localization cue. $A$ is the allowed action interface. $\Omega(a,h,E)$ is the runtime observation or feedback returned by action $a$ after history $h$ in environment $E$. $V$ is the evaluator-only final validation oracle. Thus a test file can exist in $E$ without occurring in $I$ or any permitted observation. The benchmark supplies candidate artifacts from which $E$, $I$, $L$, $\Omega$, and $V$ are constructed. We represent a reported system configuration separately as
\[
\mathcal{S}=(M,D,B,R),
\]
where $M$ is model provenance and version, $D$ is the decoding, tool-selection, and patch-selection policy, $B$ is the token, tool-call, time, and monetary budget, and $R$ records retrieval indices, random seeds, and other run controls. The permitted action interface belongs to $\mathcal{T}$. The policy that selects among those actions belongs to $\mathcal{S}$. Separating $\mathcal{T}$ from $\mathcal{S}$ avoids making every model comparison a different task by definition.

This definition intentionally does not privilege a particular evidence source. A failing test body, stack trace, retrieved historical patch, static-analysis warning, and debugger state are all evidence. They differ in provenance, timing, cost, and information content. Tests may be allowed during diagnosis, reserved for post-patch validation, or partly hidden. A test that returns feedback during a repair trajectory belongs to $\Omega$, even when a related held-out test remains in $V$. Each choice is valid for a different scientific question, but it must be reported.

A shared benchmark and final test suite provide a common outcome measure, but they do not make every component of two systems directly comparable. When $\mathcal{T}_1$ and $\mathcal{T}_2$ differ, scores answer ``Which complete system solved more instances under its stated setting?'' A claim about a model or repair method requires the relevant parts of $\mathcal{T}$ and $\mathcal{S}$ to be matched, or a controlled ablation that changes the factor of interest while holding the others fixed. Such an ablation establishes evidence only for the reported implementation and does not substitute for specifying the task contract of other results.

\begin{table}[t]
\centering\scriptsize
\caption{Manifest field layout used for each corpus record. Lists are canonicalized lexicographically. Unavailable fields are \texttt{unclear}, never inferred.}
\label{tab:manifest-example}
\begin{tabularx}{\linewidth}{p{.27\linewidth}Y}
\toprule
Field & Example value \\
\midrule
schema/version & \texttt{repair-eap/1.0} \\
evaluation protocol & \texttt{issue+checkout | agent-derived localization | read,search,build,public-test | public-test-output-iterative | F2P-held-out} \\
environment $E$ & pinned base commit, container digest, checkout without history \\
initial input/context $I$ & issue text, repository workspace, no supplied localization \\
localization $L$ & agent-derived from issue, search, and permitted test feedback \\
actions/observations $A,\Omega$ & read/search/edit/build/public-test, stdout/stderr truncated to stated limit \\
final oracle $V$ & evaluator-only FAIL\_TO\_PASS and PASS\_TO\_PASS tests \\
system $\mathcal{S}$ & model/version, decoding, seed, retrieval-index date, token/tool/time/cost caps \\
source basis & study section(s) supporting each coded value \\
\bottomrule
\end{tabularx}
\end{table}

\section{Empirical results}
\label{sec:results}

\subsection{RQ1: Variation in reported experimental settings}

The corpus contains 10 experimental settings from seven studies, each evaluating one focal system. Seven settings evaluate Defects4J variants and three evaluate SWE-bench variants. Five settings (50\%) use oracle localization, one (10\%) uses a reported localization tool, one (10\%) derives localization from issue or test information, and three (30\%) let a system derive it through repository-level actions. Six settings (60\%) explicitly describe execution or tool output being returned for subsequent repair decisions. Three (30\%) start with a repository workspace, and three (30\%) name a benchmark-harness test patch as the final validation oracle. These counts describe the studied corpus and are not estimates of field-wide practice.

\begin{table*}[t]
\centering\scriptsize
\caption{Experimental-setting results. ``Iterative feedback'' means that the study explicitly states execution or tool output is returned for subsequent repair decisions. ``Harness oracle'' denotes a benchmark evaluation test patch distinct from the agent interface.}
\label{tab:corpus-results}
\begin{tabularx}{\textwidth}{p{.16\textwidth}p{.15\textwidth}p{.16\textwidth}p{.28\textwidth}Y}
\toprule
Study / setting label & Benchmark & Localization & Initial access and repair-time feedback & Final oracle \\
\midrule
AlphaRepair / perfect & Defects4J & oracle & localized code, no explicit model feedback & project tests + assessment \\
AlphaRepair / Ochiai & Defects4J & tool-derived & ranked locations, no explicit model feedback & project tests + assessment \\
ChatRepair / line/function & Defects4J & oracle & localized function + failure information, iterative feedback & project tests + assessment \\
ThinkRepair / no feedback & Defects4J & oracle & localized function + examples, no feedback appended & project tests + assessment \\
ThinkRepair / feedback & Defects4J & oracle & localized function + examples, iterative test feedback & project tests + assessment \\
RepairAgent / default & Defects4J & agent-derived & tool-mediated project access, iterative tool output & project tests + assessment \\
Agentless / default & SWE-bench Lite & issue/test-derived & issue + repository workspace, staged feedback not established & harness test patch \\
AutoCodeRover / pass@1 & SWE-bench & agent-derived & issue + repository workspace, iterative API output & harness test patch \\
SWE-agent / default & SWE-bench & agent-derived & issue + repository workspace, iterative command output & harness test patch \\
\bottomrule
\end{tabularx}
\end{table*}

The central empirical observation is not that any one arrangement is preferable. Rather, Defects4J is used with different combinations of fault localization, repair-time feedback, and workspace access. In particular, the seven Defects4J records include five settings with oracle fault localization, one with Ochiai-based fault localization, and one in which the system performs fault localization through repository-level actions. Under the RQ2 rule, these differences in $L$, $I$, $A$, or $\Omega$ establish distinct repair tasks. Thus the same benchmark family is used to evaluate localized patch generation, repair with repeated test feedback, and repository-level diagnosis and repair.

\subsection{RQ2: Different repair tasks constructed from the same benchmark}

\paragraph{Localization changes the repair setting.} AlphaRepair reports both perfect and Ochiai-ranked fault localization and limits baseline comparisons to the same localization setting~\citep{alpharepair}. The two corpus records consequently hold the system family and reported budget constant while changing $L$, from an oracle location to tool-derived suspicious locations. A difference between these rows is evidence about the localization condition as well as patch generation. It is not a clean estimate of an underlying model difference.

\paragraph{Test execution can be repair-time evidence.} ChatRepair supplies failure information in the initial context and feeds compilation or test failures from earlier attempts into later turns~\citep{chatrepair}. Its single-line and single-function settings differ in task unit and repair budget (200 versus 100 attempts), despite sharing the same conversational mechanism. ThinkRepair separately permits a configuration with testing information appended during interaction~\citep{thinkrepair}. In these cases, the project test suite supports candidate validation and also produces diagnostic evidence. It should not be described only as a final outcome measure.

\paragraph{An action interface defines repository-level repair.} RepairAgent lets the model select among reading, search, fault-localization, patching, and testing tools, with results incorporated in its dynamic prompt~\citep{repairagent}. By contrast, Agentless reports fixed localization, repair, and validation stages without autonomous tool use~\citep{agentless}. AutoCodeRover and SWE-agent expose repository exploration through their own action interfaces~\citep{autocoderover,sweagent}. All are useful complete systems, but their result tables combine different information-acquisition and control arrangements.

\subsection{RQ3: What remains unresolved within the selected corpus}

All 10 records are coded from the reported studies. In six (60\%), the study does not establish the relevant workspace permissions sufficiently to determine whether arbitrary repository artifacts could be read. In four (40\%), it does not state a common resource bound for the coded experimental setting. These are limitations of what can be reconstructed from reported descriptions, not errors by the original authors. They also validate the conservative coding rule. Because selection required minimum detail for localization, initial evidence, and final validation, these missingness counts are conditional on the selected corpus and may understate missingness among otherwise eligible studies. An unreported permission is not evidence that the permission was absent, and a repository's stored tests are not evidence that a repair system could inspect them. Examining implementations and execution environments is a necessary extension of this study.

\section{Discussion and implications}
\label{sec:comparability}

\subsection{A common final test supports a system-level comparison}

When studies use the same benchmark instances and the same final tests, their numbers have a common outcome: how many instances each complete system repairs under its reported setting. This comparison can be useful for choosing a complete repair system. It should not be restated as evidence that one base model, prompt, or repair method is superior when the systems differ in localization, input, tool access, feedback, sampling, or budget. APR has long separated fault localization, patch generation, and validation in its repair pipeline~\citep{learningbasedsurvey}. Recent APR work likewise records input representation, fault-localization assumptions, benchmark variants, and evaluation settings~\citep{llmaprsurvey,patchgensurvey}. Our framework makes those already recognized factors explicit for each reported experimental setting.

\subsection{Localization is information}

Supplying the exact buggy method, a suspicious line, or the changed file removes a substantial part of the debugging problem. Classical APR often measures this separately through perfect versus statistical fault localization. The distinction should persist in LLM and agent evaluations. Repository-level agents may appear to avoid localization, but their search tools, issue text, and test failures provide localization evidence. A result should name its localization regime as oracle, generated, search-based, or absent.

\subsection{Validation feedback can become diagnosis feedback}

Compilation and testing are commonly described as validation. In an iterative loop, however, their output is also information used to choose the next edit. The difference between seeing only a final evaluator verdict and seeing every failing assertion, stack trace, sanitizer report, and coverage trace can be decisive. Report the feedback granularity, test identities accessible to the agent, maximum validation rounds, and whether the feedback originates from public or hidden tests.

\subsection{Sampling and budget change the measured capability}

Pass@$k$ measures success under $k$ sampled candidates, not the quality of one repair trajectory. It is sensitive to temperature, candidate count, selection rule, and validation budget. Agentic results add tool calls, context growth, runtime, and monetary cost. A comparison between a single-shot model and an agent allowed dozens of test-driven attempts can be useful, but should be framed as a system-level trade-off rather than a base-model comparison. RepairAgent's explicit accounting of time, tokens, and cost is a positive example~\citep{repairagent}.

\subsection{Contamination and temporal validity}

Public historical bugs, reference patches, issue discussions, and benchmark tests may occur in model pretraining or retrieval corpora. This does not make an experiment useless, but it changes its interpretation. Temporal splits, post-training tasks, decontamination searches, held-out repositories, and transparent model release dates improve evidence. The risk is especially acute for small, long-lived benchmarks and for unrestricted web-enabled agents. Comparative reports and leaderboards should avoid interpreting a number as pure reasoning ability without considering exposure.

\begin{table}[t]
\centering\small
\caption{Minimum fields needed to interpret a repair result.}
\label{tab:manifest}
\begin{tabularx}{\linewidth}{p{.28\linewidth}Y}
\toprule
Field & Reportable value \\
\midrule
Evaluation protocol & task unit/language, environment, initial input/context, localization source \\
Action and observation & allowed tools, commands, network/history access, output truncation, feedback rounds \\
Final oracle & tests used only after repair, PoC, independent correctness assessment \\
System configuration & model/version, decoding, patch selection, samples, tokens, calls, time, monetary cost \\
Run controls & commit/environment/container, seeds, logs, retrieval index/date \\
Outcome label & test-adequate, independently validated, human assessed, or unclear \\
\bottomrule
\end{tabularx}
\end{table}

\subsection{Reporting implications}

\subsubsection{Publish an evidence manifest}

Every repair experiment should publish the fields in Table~\ref{tab:manifest}, preferably in machine-readable form for each reported experimental setting. A concise manifest could be attached to every headline result. For example, it could state \emph{initial: issue + repository. Actions: read/search/build/public tests. Evaluator-only: FAIL\_TO\_PASS tests. Localization: agent-derived. Budget: 30 tool calls and 2M tokens}. This adds little burden while preventing many invalid comparisons.

The manifest should serialize $\mathcal{T}$ and $\mathcal{S}$ separately, record a schema version plus benchmark and environment identifiers, sort set-valued fields canonically, and preserve a source location for each value. It should distinguish permitted actions from actions exercised in a particular trace. Recording both supports capability-level comparison without confusing it with a single successful trajectory. Missing information must remain \texttt{unclear}. It must not be reconstructed from a dataset layout or a default tool setting.

\section{Conclusion}
\label{sec:conclusion}

LLM-based and agentic repair are expanding APR's capabilities and its experimental degrees of freedom. In the studied corpus of 10 experimental settings, studies using one benchmark family employ oracle, tool-derived, and agent-derived localization. Six settings feed execution or tool observations into subsequent repair decisions. Reported descriptions frequently leave workspace permissions or a common budget unresolved. These observations explain why a statement that two studies use the same benchmark is insufficient to identify the same repair task. When systems use the same benchmark instances and final tests, their scores can be compared as overall performance of the complete systems under their reported settings. If fault localization, input, feedback, tools, sampling, or budget differs, the score difference alone does not establish that one model or repair method is better. A controlled ablation can establish the effect of one factor within a reported implementation, while a common specification is needed to interpret independently reported tasks. The framework makes those conditions explicit and provides a basis for future work that examines implementations and execution environments.

\bibliographystyle{plainnat}
\bibliography{references}
\end{document}